%% file: selfcons.tex
\documentstyle[epsf,12pt,fleqn]{article}

\def\eqref#1{(\ref{#1})}
\renewcommand{\matrix}[4]{\left( \begin{array}{cc}
        #1 & #2 \\ #3 & #4 \end{array}\right)}

\newcommand{\boldtau}{\mbox{\boldmath$\tau$\unboldmath}}

\begin{document}
\bibliographystyle{unsrt}

\font\twelvemb=cmmib10 scaled \magstep1
\font\tenmb=cmmib10
\font\ninemb=cmmib9
\font\sevenmb=cmmib7
\font\sixmb=cmmib6
\font\fivemb=cmmib5
\font\twelvesyb=cmbsy10 scaled \magstep1
\font\tensyb=cmbsy10
\font\sstwelve=cmss12
\font\ssnine=cmss9
\font\sseight=cmss8
\textfont9=\twelvemb
\scriptfont9=\tenmb
\scriptscriptfont9=\sevenmb
\textfont10=\twelvesyb
\def\bm{\fam9}
\def\bms{\fam10}

\newcommand{\DS}{\displaystyle}
\newcommand{\TS}{\textstyle}
\newcommand{\SS}{\scriptstyle}
\newcommand{\SSS}{\scriptscriptstyle}

\newcommand\zZtwelve{\hbox{\sstwelve Z\hskip -4.5pt Z}}
\newcommand\zZnine{\hbox{\ssnine Z\hskip -3.9pt Z}}
\newcommand\zZeight{\hbox{\sseight Z\hskip -3.7pt Z}}
\newcommand\zZ{\mathchoice{\zZten}{\zZten}{\zZeight}{\zZeight}}
\newcommand\ZZ{\mathchoice{\zZtwelve}{\zZtwelve}{\zZnine}{\zZeight}}

\mathchardef\sigma="711B
\mathchardef\tau="711C
\mathchardef\omega="7121
\mathchardef\nabla="7272

\newcommand{\e}{\epsilon}
\newcommand{\eps}{\epsilon}
\newcommand{\ee}{\varepsilon}
\newcommand{\vp}{\varphi}
\newcommand{\vphi}{\varphi}
\newcommand{\cphi}{\Phi}
\let\oldvrho=\varrho
\newcommand{\vrho}{{\raise 2pt\hbox{$\oldvrho$}}}
\let\oldchi=\chi
\renewcommand{\chi}{{\raise 2pt\hbox{$\oldchi$}}}
\let\oldxi=\xi
\renewcommand{\xi}{{\raise 2pt\hbox{$\oldxi$}}}
\let\oldzeta=\zeta
\renewcommand{\zeta}{{\raise 2pt\hbox{$\oldzeta$}}}

\newcommand{\la}[1]{\label{#1}}
\newcommand{\ur}[1]{(\ref{#1})}
\newcommand{\ra}[1]{(\ref{#1})}
\newcommand{\urs}[2]{(\ref{#1},~\ref{#2})}
\newcommand{\eq}[1]{eq.~(\ref{#1})}
\newcommand{\eqs}[2]{eqs.~(\ref{#1},~\ref{#2})}
\newcommand{\eqss}[3]{eqs.~(\ref{#1},~\ref{#2},~\ref{#3})}
\newcommand{\eqsss}[2]{eqs.~(\ref{#1}--\ref{#2})}
\newcommand{\Eq}[1]{Eq.~(\ref{#1})}
\newcommand{\Eqs}[2]{Eqs.~(\ref{#1},~\ref{#2})}
\newcommand{\Eqss}[3]{Eqs.~(\ref{#1},~\ref{#2},~\ref{#3})}
\newcommand{\Eqsss}[2]{Eqs.~(\ref{#1}--\ref{#2})}
\newcommand{\fig}[1]{Fig.~\ref{#1}}
\newcommand{\figs}[2]{Figs.~\ref{#1},\ref{#2}}
\newcommand{\figss}[3]{Figs.~\ref{#1},\ref{#2},\ref{#3}}
\newcommand{\beq}{\begin{equation}}
\newcommand{\eeq}{\end{equation}}

\newcommand{\doublet}[3]{\:\left(\begin{array}{c} #1 \\#2
            \end{array} \right)_{#3}}
\newcommand{\vect}[2]{\:\left(\begin{array}{c} #1 \\#2
            \end{array} \right)}
\newcommand{\vectt}[3]{\:\left(\begin{array}{c} #1 \\#2 \\ #3
            \end{array} \right)}
\newcommand{\vectf}[4]{\:\left(\begin{array}{c} #1 \\#2 \\#3\\#4
            \end{array} \right)}
\newcommand{\matr}[4]{\left(\begin{array}{cc}
                   #1 &#2 \\
                   #3 &#4 \end{array} \right)}
\newcommand{\fracsm}[2]{{\textstyle\frac{#1}{#2}}}

\newcommand{\D}{{\cal D}}
\newcommand{\K}{{\cal K}}
\newcommand{\NC}{N_{\rm CS}}
\newcommand{\Ncs}{N_{\rm CS}}
\newcommand{\SU}{$SU(2)~$}
\newcommand{\Pmax}{P_{max}}
\newcommand{\tr}{\,{\rm tr}\,}
\newcommand{\Tr}{\,{\rm Tr}\,}
\newcommand{\ldef}{=}
\newcommand{\rdef}{=}
\newcommand{\simlt}{\stackrel{<}{{}_\sim}}
\newcommand{\simgt}{\stackrel{>}{{}_\sim}}

\newcommand{\nuH}{\nu_{\SSS H}}
\newcommand{\nuF}{\nu_{\SSS F}}
\newcommand{\nuf}{\nu_{\SSS f}}
\newcommand{\nut}{\nu_{\SSS t}}
\newcommand{\nuHold}{\nu_{{\SSS H}\,{\rm old}}}

\newcommand{\op}[1]{{\bf \hat{#1}}}
\newcommand{\opr}[1]{{\rm \hat{#1}}}
\newcommand{\bra}[1]{\langle#1\vert}
\newcommand{\ket}[1]{\vert#1\rangle}
\newcommand{\lsim}{\mathrel{\lower 2pt\hbox{$\stackrel{<}{\SS\sim}$}}}
\newcommand{\gsim}{\mathrel{\lower 2pt\hbox{$\stackrel{>}{\SS\sim}$}}}




{\thispagestyle{empty}
\begin{center}
{\Large\bf
Energy of electroweak sphalerons in $1$-loop
selfconsistent calculations\\}
\vspace{27 pt}
{\large\bf
Wolfram Schroers$^*$,\footnote{
\noindent
Wolfram.Schroers@tp2.ruhr-uni-bochum.de}\\
Ingo B\"ornig$^{*}$,\footnote{
\noindent
Ingo.Boernig@tp2.ruhr-uni-bochum.de}
Christian Schulzky$^{**}$,\footnote{
\noindent
c.Schulzky@physik.tu-chemnitz.de}\\
and Klaus Goeke$^{*}$\footnote{
\noindent
goeke@hadron.tp2.ruhr-uni-bochum.de}}\\
\vspace{20 pt}
{\small\it $^{*}$
Institut f\"ur Theor.~Physik I\hskip-1.5pt I, Ruhr-Universit\"at
Bochum, D-44780 Bochum, Germany}\par
{\small\it $^{**}$
Institut f\"{u}r Physik, TU-Chemnitz, D-09107 Chemnitz, Germany}
\end{center}
\vspace{30 pt}
\abstract{We calculate the energy of electroweak sphalerons including
one-loop fermionic corrections. This calculation has previously been
done by adding the correction to the tree-level sphaleron and
interpreting the resulting energy as the $1$-loop energy. However, in
this paper we calculate the energy by doing a full renormalisation
of the parameters on $1$-loop-level and redetermining the sphaleron
configuration using the full energy functional. When comparing the
final result with the tree-level solution we find that the $1$-loop
calculation will only cause small deviations of the sphaleron energy.
}}
\newpage

\input chap1

\input chap2
\input chap3

\input chap4

\newpage
\appendix
\input selfconsa
\input nondeg

\newpage
\bibliography{literature}

\end{document}

%% file: chap1.tex

\section{Introduction}

In the study of non-abelian gauge theories it has been found that 
the non-trivial structure of the gauge fields can lead to baryon 
number violating processes due to the anomaly of baryon and lepton 
currents (see \cite{thooft}). This fascinating property of 
Yang-Mills theory was discovered by Faddeev \cite{faddeev} and
by Jackiw and Rebbi \cite{jackreb} who showed that different 
field configurations
of the gauge-fields exist that are topologically distinct but are 
physically all vacuum configurations. Thus the field energy of these
configurations will be zero. These configurations are numbered by
a certain functional of the fields, the Chern-Simons number 
$N_{\rm CS}$ (which is, mathematically speaking, the winding
number). They can be 
continuously deformed into each other but this process will violate
the boundary condition that the energy is zero. However, there exists
a way of deforming the fields that will make the required energy as
low as possible. The peak of this path is known as the sphaleron and
its energy is of vital interest for the whole process.
The energy of the sphaleron is of the order of $m_W/\alpha$
where $m_W$ is the mass of the $W$ boson and $\alpha$ the $SU(2)$
coupling constant.

The classical sphaleron energy (the tree-level result) has been known
for some time already. Naturally the question arises, how accurate the
tree-level result is. Quantum corrections to the sphaleron have been
computed by Bockharev and Shaposhnikov; they have included
boson fluctuations through an effective potential of the Higgs field
(see \cite{bok,shap}). A direct computation of the bosonic 
determinant over nonzero modes has been performed by
Carson and McLerran \cite{carson,carson2} and Baacke and Junker 
\cite{baacke}.

In \cite{DPSSG} for the first time fermionic quantum 
corrections both at zero and at finite temperature
were examined and their influence turned out to be 
significant. However, these calculations performed the 
renormalisation only approximately and they did not take
into account that the gauge fields might look different on $1$-loop
level than on tree-level. 
The generalisation of these result to the case 
including bosonic contributions was done in \cite{sphaleronpaper}.
However, the conclusions drawn there were based on the above-mentioned
simplifications in the renormalization and differ from the 
results given by Moore in
\cite{moorepaper} and Shaposhnikov \cite{shaposhnikov}.

The computation of the one-loop sphaleron configuration by using a 
self-consistent approach is the primary aim of our
paper. We further perform a correct renormalisation of all parameters
of the theory and discuss the effects on the results given in
\cite{sphaleronpaper}.

In section 2 we discuss the background of sphaleron configurations and
fermionic fluctuations. Section 3 discusses the methods and results
of the self-consistent calculation. In the last chapter we draw the
conclusions and relate this work to other papers.

%% file: chap2.tex

\section{Sphalerons and fermionic fluctuations}\label{chap2}

The particle model under examination is the minimal standard model
with one Higgs doublet. In \cite{noweinberg} it has been analysed,
how large the influence of the $U(1)$-sector to the sphaleron energy
is. As its influence is only of the order of $\simeq 1\%$ it seems to
be justified to neglect the $U(1)$-sector in the model 
(by setting the Weinberg angle $\vartheta_W=0$) and work only
with the $SU(2)$-symmetric Yang-Mills theory with a single Higgs
doublet. The Lagrangian is thus (for one fermion species)
\begin{eqnarray}
{\cal L} = & \displaystyle\overline{\psi}_L i \gamma^\mu D_\mu\psi_L
                + \overline{\psi}_R i \gamma^\mu\partial_\mu\psi_R
                - \overline{\psi}_L M\psi_R
                - \overline{\psi}_R M^\dagger\psi_L\nonumber \\
           & \displaystyle - {1\over 4g^2}F^a_{\mu\nu}F^{a\mu\nu}
                + \left(D_\mu\Phi\right)^\dagger\left(D^\mu\Phi\right)
                - \frac{\lambda^2}{2}\left(\Phi^\dagger\Phi
                - {v^2\over 2}\right)^2
\label{lagrange}
\end{eqnarray}
where we introduced the covariant derivative 
$D_\mu=\partial_\mu-iA_\mu$ with $A_\mu=
{1\over 2}A^a_\mu\tau^a$ and the field strength tensor $F_{\mu\nu}={1\over 2}
F_{\mu\nu}^a\tau^a$, where $F_{\mu\nu}^a=\partial_\mu A^a_\nu-\partial_\nu
A_\mu^a+\epsilon^{abc}A_\mu^b A_\nu^c$. The mass matrix $M$ consists of
the components of the Higgs field $\Phi={\Phi^+ \choose \Phi^0}$ and the
Yukawa couplings $h_u$ and $h_d$,
\begin{equation}
M = \left( \begin{array}{cc}
        h_u\Phi^{0*} & h_d\Phi^+ \\
        -h_u\Phi^{+*} & h_d\Phi^0 \\
\end{array}\right).
\end{equation}
The $SU(2)$ fermion doublets are defined by
\[
\psi_L={1\over 2}(1-\gamma_5)\psi={\psi_L^u\choose\psi_L^d},
\]
\[
\psi_R={1\over 2}(1+\gamma_5)\psi={\psi_R^u\choose\psi_R^d}.
\]
Masses are generated by the non-vanishing vacuum expectation value
of the Higgs field
$\langle 0\vert\Phi\vert 0\rangle=\frac{v}{\sqrt{2}}{0\choose 1}$. 
This mechanism yields on tree-level
$m_W={g v\over 2}$, $m_{u,d}={h_{u,d} v\over\sqrt{2}}$ and $m_H=\lambda v$.
The renormalised masses on $1$-loop level are explicitely given in
appendix \ref{appendixmass} and will be needed later for obtaining the
selfconsistent solution.

In order to be consistent with previous works we choose to scale the
quantities in the following way:
\[
x^\mu\to m_W^{-1}x^\mu,\qquad
A_\mu^a\to m_WA_\mu^a, \qquad
\Phi\to{m_W\over\sqrt{2}g}\Phi.
\]
Furthermore we use the following representation of the Dirac matrices
\begin{equation}
\gamma^0=\matrix{0}{1}{1}{0},\quad 
\gamma^i=\matrix{0}{\sigma_i}{-\sigma_i}{0}\quad \mbox{and} \quad
\gamma^5=\matrix{-1}{0}{0}{1}.
\end{equation}
which allows us to reduce the fermion fields to two components:
\[
\psi_L^{u,d}\to m_W^{3/2}{\psi_L^{u,d}\choose 0},\qquad
\psi_R^{u,d}\to m_W^{3/2}{0\choose\psi_R^{u,d}}.
\]
With these replacements the new Lagrangian becomes
\begin{eqnarray}
{\cal L}&=&m_W^4\Bigg(\displaystyle
        i\psi_L^\dagger(D_0-\sigma_iD_i)\psi_L
        +i\psi_R^\dagger(\partial_0+\sigma_i\partial_i)\psi_R
        -\psi_L^\dagger M\psi_R - \psi_R^\dagger M^\dagger\psi_L
        \nonumber\\
        &&\displaystyle+{1\over g^2}\left(-{1\over 4}F^a_{\mu\nu}
        F_a^{\mu\nu}+{1\over 2}(D_\mu\Phi)^\dagger(D^\mu\Phi)
        -{1\over 32}\nu_H^2\left(\Phi^\dagger\Phi-4\right)^2
        \right)\Bigg),
\label{lagrange2}
\end{eqnarray}
with the mass matrix
\begin{equation}
M={1\over 2}{\matrix{\nu_u\Phi^{0*}}{\nu_d\Phi^+}
        {-\nu_u\Phi^{+*}}{\nu_d\Phi^0}}.
\label{massmat}
\end{equation}
The masses have to be taken from experiment. However, when comparing
the masses of the different fermions, one finds that the top quark
largely dominates the whole fermionic mass spectrum. Its energy of
$\simeq 175$GeV is much larger than that of any other quark, so the
other fermion doublets can be considered as massless. In \cite{DPSSG}
the authors find that the influence of massless fermion doublets is
negligibly small when compared to a massive doublet, so only the
top/bottom doublet needs to be considered. However, when applying
the usual framework to compute the energy spectrum, only doublets that
are degenerate in mass can be computed. Consequently the top/bottom
doublet was approximated by $1.5$ fermionic doublets with a 
degenerate mass of $m_T$ and a corresponding bare value of
$\nu_F=\nu_u=\nu_d$. To estimate the effect of nondegenerate 
fermion masses we
developed a perturbative method, which can be used to 
justify this approach. In
Appendix \ref{nondeg} and \cite{diplomingo} we present these results.
The parameters in this work are chosen to be
\cite{topmass}
\[
g=0.67,\quad m_W=83\mbox{GeV},\quad m_T=175\mbox{GeV}.
\]
The Higgs mass will be considered as a free parameter. One aim of this
work is to express the sphaleron energy as a function of the Higgs
mass.
The relation of the bare parameters $\nu_F$ and $\nu_H$ to these 
physical parameters is discussed below.

Now we will focus our attention to static solutions of the gauge and
Higgs-fields. In this case it is useful to use the temporal gauge with
$A_0\equiv 0$. In this gauge \eqref{lagrange2} yields the following
energy functional:
\begin{equation}
E_{\rm class}={m_W\over g^2}\int d^3r\left({1\over 4}F^a_{ij}F_a^{ij}+
        {1\over 2}\left(D_i\Phi\right)^\dagger\left(D_i\Phi\right)+
        {1\over 32}\nu_H^2\left(\Phi^\dagger\Phi -4\right)^2\right).
\label{classerg}
\end{equation}
Configurations with $A_i^a=0$ and $\Phi^\dagger\Phi=4$ have 
$E_{\rm class}=0$ which means they are vacuum configurations. Due to the
transformation properties of \eqref{lagrange2} one can transform this
configuration to $A_i'=iU(x)\partial_iU^\dagger(x)$ (with $U(x)\in 
SU(2)$) where the new configuration must still have $E_{\rm class}=0$,
due to gauge invariance.

The field configurations can furthermore be characterised by the following
functional (called Chern-Simons-number):
\begin{equation}
N_{\rm CS}={1\over 16\pi^2}\int d^3r\epsilon^{ijk}\left(
        A_i^a\partial_jA^a_k+{1\over 3}\epsilon^{abc}
        A_a^iA_b^jA_c^k\right).
\label{ncs}
\end{equation}
One can show that for configurations of the form 
$A_i=iU(x)\partial_iU^\dagger(x)$ the Chern-Simons-number is an
integer value
$N_{\rm CS}\in \ZZ$. If $N_{\rm CS}$ is not an integer then 
\eqref{classerg} will yield an energy larger than $0$.
Thus it is impossible to deform vacuum configurations with different $N_{\rm
  CS}$ continously into each other without violating the boundary condition
$E_{\rm class}=0$. 

Of special interest is the path of lowest energy going from
$N_{\rm CS}=0$ to $N_{\rm CS}=1$. It can be obtained by minimising the
energy functional \eqref{classerg} for a given and fixed $N_{\rm CS}$.
This path has been constructed in \cite{aky}.
From symmetry arguments the point at $N_{\rm CS}=1/2$ is the highest
point and is called {\sl sphaleron}. It is a saddle point in the space
of Higgs-  and gauge fields. There is only one negative eigenmode, which
is lying in the direction of changing $N_{\rm CS}$. As this point already
has a high symmetry in the space of Higgs- and gauge-fields it appears
logical to choose a spherically symmetric ansatz for the spatial dependence of
the fields. This ansatz is called {\sl hedgehog} and has been used
in a variety of works on this subject 
\cite{aky,DPSSG,diplomingo,docpeter}. The Higgs- and gauge-fields are
represented by $5$ functions $A(r), B(r), C(r), G(r)$ and $H(r)$ in the
following way:
\begin{eqnarray}
A_i^a({\bf r})&=&\epsilon_{aij}n_j{1-A(r)\over r}+
        (\delta_{ai}-n_an_i){B(r)\over r}+
        n_an_i{C(r)\over r}, \nonumber \\
\Phi({\bf r})&=&2\left( H(r)+iG(r){\bf n\cdot\bm\boldtau}\right)
        {0\choose 1}.
\label{hedgehog}
\end{eqnarray}
In this ansatz the energy \eqref{classerg} has the following form:
\begin{eqnarray}
E_{\rm class}&=&\displaystyle{4\pi m_W\over g^2}\int_0^R dr\Biggl(
        \left(A^\prime+{BC\over r}\right)^2+
        \left(B^\prime-{AC\over r}\right)^2\nonumber\\
        &&\displaystyle+{1\over 2}\left({A^2+B^2-1\over r}\right)^2
        +2r^2\left(H^{\prime 2}+G^{\prime 2}\right)
        \nonumber\\
        &&\displaystyle +2r\left( H^\prime G-G^\prime H\right)C-
        4BGH+2A\left(G^2-H^2\right)\nonumber \\
        &&\displaystyle +\left(1+A^2+B^2+{1\over 2}C^2\right)
        \left(G^2+H^2\right)+{\nu_H^2\over 2}r^2
        \left( G^2+H^2-1\right)^2\Biggr).
\label{classerg2}
\end{eqnarray}
For $N_{\rm CS}$ one obtains the following expression
\begin{equation}
N_{\rm CS}={1\over 2\pi}\int_0^\infty dr\left(
{(A^2+B^2-1)C\over r}+(A^\prime B-B^\prime A)+B^\prime\right).
\label{ncs2}
\end{equation}
The fermionic fluctuations have been investigated in \cite{DPSSG}.
Here we use only the expression for the renormalised fermionic
field energy, which is a function of some numerical parameters
and the physical quantities $\nu_F$, $\nu_H$ and $m_W$. An additional
parameter has to be introduced, the renormalisation point 
$\nu_{\rm ren}$. Of course, the final results should be independent
of $\nu_{\rm ren}$. Note that
for the $\nu_F$ and $\nu_H$ one has to use the renormalised
quantities on the $1$-loop-level (see appendix \ref{appendixmass}).

The numerical parameters, as introduced in \cite{DPSSG} are the
box size $R$ and the cut-off $\Lambda$. Thus the whole fermionic
energy at zero temperature becomes
\begin{equation}
E_{\rm ferm}^{\rm ren}(\Lambda,R,\nu_F,\nu_H,\nu_{\rm ren})
=E_{\rm sea}(\dots)-E_{\rm div}(\dots),
\label{fermerg}
\end{equation}
where the sea energy and the divergent energy are
\begin{equation}
        E_{\rm sea}(\Lambda)={m_{W}\over 
        4\sqrt{\pi}}\int_{\Lambda^{-2}}^{\infty} {dt\over t^{3/2}} {\rm 
        Tr}\left( e^{-t{\cal H}^2}-e^{-t{{\cal H}^{(0)}}^2}\right)
        \label{eseacutoff},
\end{equation}
\begin{eqnarray}
E_{\rm div}(\Lambda)&=&%
\displaystyle{m_{W}\over 32\pi^2}\int d^3r\left(\nu_{\rm ren}^2- 
\Lambda^2\right)\nu_{F}^2\left(\Phi^\dagger\Phi-4\right)
\nonumber \\ &&\displaystyle 
+\ln{\Lambda^2\over\nu_{\rm ren}^2}\Biggl({1\over 6}\left( 
F^a_{ij}\right)^2+\nu_{F}^2\left( D_{i}\Phi\right)^\dagger\left( 
D_{i}\Phi\right)\nonumber\\
&&\displaystyle \qquad +{1\over 4}\nu_{F}^4\left(\left( 
\Phi^\dagger\Phi-4\right)^2+8\left(\Phi^\dagger\Phi-4\right)\right)\Biggr).
\label{ediv}
\end{eqnarray}
The fermionic Hamiltonian is defined to be
\begin{equation}
{\cal H}=\matrix{i\sigma_iD_i}{M}{M^\dagger}{-i\sigma_i\partial_i}.
\label{hamilton}
\end{equation}
The task of evaluating the energy contribution \eqref{eseacutoff}
by diagonalising \eqref{hamilton} in an appropriate basis has been 
described thoroughly in \cite{DPSSG} and shall not be repeated here.

Before we further investigate the $1$-loop energy expressions we shall
examine the parameters of the Lagrangian on $1$-loop, $\nu_F$, $\nu_H$
and $\nu_{\rm ren}$. On tree level, the relations between these parameters
and the physical particle masses are trivial and have been given above.
But on $1$-loop-level these relations become extremely complicated. One
gets a set of three equations depending on the three parameters. One
equation can be eliminated to define the mass scale of the system. It
appears natural to choose the equation for $m_W$. Thus, finally 
\begin{eqnarray}
m_H&=&m_H(\nu_{\rm ren},\nu_F,\nu_H),\nonumber \\
m_F&=&m_F(\nu_{\rm ren},\nu_F,\nu_H).
\label{massrenorm}
\end{eqnarray}
These system of equations is overdetermined and we have the freedom to
choose one parameter arbitrarily. Usually one takes $\nu_{\rm ren}$ as
a free parameter and examines if the final results are independent of
$\nu_{\rm ren}$. The full set of equations \eqref{massrenorm} is given
in appendix \ref{appendixmass}.

Furthermore we need the vacuum expectation value of the Higgs field
in the $1$-loop case. To obtain this value we have to examine the value
of the fermionic energy in the vacuum which is given by ($\Gamma$ is
Euler's constant):
\begin{eqnarray}
\displaystyle E_{\rm ferm}^{\rm ren}[F_{ij}^a= 0, 
\Phi={\rm const}]&=&\displaystyle
{m_{W}N_{c}\over 32\pi^2}\int d^3r\Biggl(-\nu_{F}^2\nu_{\rm ren}^2 
\Phi^\dagger\Phi\nonumber\\
&&+\displaystyle {\nu_{F}^2\left(\Phi^\dagger\Phi\right)^2\over 
8}\left({3\over 2}-\Gamma-\ln\left({\nu_{F}^2\Phi^\dagger\Phi\over 
4\nu_{\rm ren}^2}\right)\right)\Biggr).
\label{vacferm}
\end{eqnarray}
By performing $\frac{\delta E_{\rm ferm}^{\rm ren}}{\delta\left(
\Phi^\dagger\Phi\right)}=0$ one gets the equation for $v_1$:
\begin{equation}
        v_{1}^2-4=-{N_{c}g^2\over 2\nu_{H}^2\pi^2}\left( -\nu_{F}^2\nu_{\rm 
        ren}^2+{\nu_{F}^4v_{1}^2\over 4}\left( 
        1-\Gamma-\ln\left({\nu_{F}^2v_{1}^2\over 4\nu_{\rm 
        ren}^2}\right)\right)\right).
        \label{v1det}
\end{equation}
This equation is to be solved numerically for given $\nu_F$, $\nu_H$ and
$\nu_{\rm ren}$. It should be noted that there is an apparent dependence
of $v_1$ on $\nu_{\rm ren}$, but by renormalising the mass-scale (see above)
this dependence is removed. Furthermore it should be noted that massless
fermions do not influence the value of $v_1$; since the energy in 
\eqref{vacferm} is $\propto \nu_F^2$ the influence of massless fermions
is $0$.

With the help of the expression for the classical energy (the tree
level functional) \eqref{classerg} and the $1$-loop fluctuations in
\eqref{fermerg} the correct expression for the $1$-loop energy functional
can be obtained (now we have to use the $1$-loop values for $\nu_F$, 
$\nu_H$ and $v_1$):
\begin{eqnarray}
        E[A,\Phi]&=&
        \displaystyle E_{\rm class}+E_{\rm ferm}^{\rm ren}-\left( E_{\rm 
        class}+E_{\rm
        ferm}^{\rm ren}\right)\vert_{\Phi^\dagger\Phi=v_{1}^2}\nonumber\\
        &=&\displaystyle{m_{W}\over g^2}\int d^3r\left( {1\over 4}\left( 
        F_{ij}^a\right)^2+{1\over 2}\left(D_{i}\Phi\right)^\dagger\left( 
        D_{i}\Phi\right)+{\nu_{H}^2\over 32}\left(\Phi^\dagger\Phi-v_{1}^2 
        \right)^2\right)\nonumber\\
        &&\displaystyle +E_{\rm ferm}^{\rm ren}-E_{\rm ferm}^{\rm ren}
        \vert_{\Phi^\dagger\Phi^2=v_{1}^2}+ {m_{W}\nu_{H}^2\over 
        16g^2}(v_{1}^2-4) \int d^3r\left(\Phi^\dagger\Phi -v_{1}^2\right).
        \label{esigmaold}
\end{eqnarray}
After the following replacement
\[
\Phi\to\Phi\cdot{v_1\over 2},\qquad
A\to A\cdot{v_1\over 2},\qquad
r\to r\cdot{2\over v_1},\qquad
\nu_{\rm ren}\to\nu_{\rm ren}\cdot{2\over v_1},
\]
one finally gets the following form of \eqref{esigmaold}
\begin{equation}
        E[A,\Phi]={v_{1}\over 2}\left(E_{\rm class}+
        E_{\rm ferm}^{\rm ren}
        \vert_{\nu_{\rm ren}{2\over v_{1}}}
        +{m_{W}\nu_{H}^2\over 4g^2}\left(1-{4\over 
        v_{1}^2}\right)\int d^3r\left(\Phi^\dagger\Phi -4\right)
        \right).
        \label{esigma}
\end{equation}
The task of finding the sphaleron configuration now requires minimising
the functional \eqref{esigma} at a given value for $N_{\rm CS}$. The
question is if this value is bound to be $N_{\rm CS}=1/2$ as it was
in the classical case. Clearly the answer is no; since the whole path
is not symmetric (if the vacuum at $N_{\rm CS}=0$ has an energy of
$E=0$ then the vacuum at $N_{\rm CS}=1$ will have the energy of the
created valence fermions which is clearly $>0$) we would expect the
highest point to lie at $N_{\rm CS}>1/2$, but still very close to it.
In principle, the difference in energy should be less than the energy
of the valence fermions which is $\simeq 1\%$ of the sphaleron
energy. So it appears to be justified to take the minimal value of
\eqref{esigma} at $N_{\rm CS}=1/2$ as an approximation for the
sphaleron energy without introducing larger errors.

%% file: chap3.tex

\section{Numerical results}

Now we wish to minimise the functional \eqref{esigma} at the fixed
value for $N_{\rm CS}=1/2$. In the $1$-loop case with the very
complicated expression \eqref{fermerg} this is a very complex task
especially since the condition for $N_{\rm CS}$ cannot be cast
into some explicit form for the hedgehog-fields. The problem arises from the
fact that the sphaleron configuration is rather a saddle point than a
minimum. So we have to modify the functional in a way that allows us to apply
some minimisation procedure.

The most natural modification involving the boundary condition
$N_{\rm CS}=1/2$ is to add a term of the form
\[
a\left(N_{\rm CS}-1/2\right)^b
\]
with some parameters $a$ and $b$. It has turned out that the choices
of $b=2$ (higher values would make the minimum too large) and 
$a=1000m_W$ (which is one magnitude higher than the sphaleron energy)
give good results. So finally one ends up with a new functional
\begin{equation}
E_\Sigma=E_X[A,\Phi]+a\left(N_{\rm CS}[A,\Phi]-1/2\right)^2.
\end{equation}
$E_X$ may either be \eqref{classerg} or \eqref{esigma} since the
procedure may also be applied to the classical case.

At this moment we still have a gauge freedom we can use to fix the
shape of the $C$-field appearing in the hedgehog ansatz in
\eqref{hedgehog}. We choose the following form which turned out to
be useful for our purposes
\[
\bar{C}(x)=-280\pi\left({x\over s}%
\right)^4\left({x\over s}-1\right)^4.
\]
The quantity $s$ has been set to $s=3m_W$ which is the size of the
other hedgehog fields, too.

In order to do the actual minimisation, the fields $A$, $B$, $G$ and
$H$ have been discretised. For the minimisation procedure, the Powell
method (see e.g. \cite{numrecipes})
was used and the fields are allowed to be modified in the range
from $r=0$ to $r=R_{\rm sphal}$. $R_{\rm sphal}$ is thus the maximum
size of the sphaleron that can be found with this method. Obviously
$R_{\rm sphal}$ has to be chosen such that 
\begin{equation}
R_{\rm sphal}\leq R,
\label{consist}
\end{equation}
where $R$ is the radius of the box for the Kahana-Ripka basis 
(see \cite{DPSSG}). The problem is that the symmetries of
the Lagrangian \eqref{lagrange2} are destroyed when one comes close
to the point $r=R$; so a minimisation procedure that searches for
a sphaleron in the free space (where $R\rightarrow\infty$) will 
encounter instabilities when it tries to manipulate the fields 
too close to the boundary. This problem is very severe if one uses
a minimisation procedure which involves computing the functional 
derivatives of the fields (see \cite{weisspaper}). However, since
our method does not involve derivatives it should be stable and
reliable if we ensure that \eqref{consist} is satisfied.

Before performing this examination the lattice spacing should
be examined first; the number of lattice points of the different 
fields are presented in table \ref{hugecalcs}.
The bare parameters of the Lagrangian have been chosen to be
$\nu_H=0.8$, $\nu_F=2.1$ and $\nu_{\rm ren}=2.0$. We find that
the first row in the table is fine enough to represent the
fields.

\begin{table}
\begin{center}
\begin{tabular}{|c|c|c|c||c|}
\hline
$n_a$ & $n_b$ & $n_g$ & $n_h$ & $E_\Sigma/$GeV \\ \hline\hline
$31$ & $21$ & $21$ & $35$ & $12353$ \\ \hline
$59$ & $43$ & $43$ & $59$ & $12356$ \\ \hline
$149$ & $109$ & $109$ & $149$ & $12356$ \\ \hline
\end{tabular}
\end{center}
\caption{The dependence of the self-consistent solution on the 
number of lattice points.
\label{hugecalcs}}
\end{table}

In order to determine an appropriate value for $R_{\rm sphal}$, we
examined the energy functional \eqref{esigma}.
We found that the total energy does not depend on the choice of $R_{\rm
  sphal}/R$, and it is convenient to use $R_{\rm sphal}=%
0.70R$ as a good compromise between accuracy and computational 
effort of the problem.

Now we can examine three different solutions: the tree-level sphaleron
(the sphaleron solution of \eqref{classerg} considering only the classical
energy of the boson field), the one-loop sphaleron (the field configuration
of the tree-level sphaleron evaluated with \eqref{esigma}) and the
selfconsistent sphaleron (the sphaleron solution of \eqref{esigma}).

By computing the energy of these solutions and comparing the deviations
one can make a statement about the significance of self-consistency in
the one-loop approximation of the sphaleron. The actual results for
different values of the Higgs mass are listed in table \ref{selfconstab}.

\begin{table}
\begin{center}
\begin{tabular}{|c|c|c|}
\hline
$m_H/$GeV & $E_{\Sigma,0}/$GeV & $E_{\Sigma,mini}/$GeV \\ \hline
$83$ & $8804$ & $8746$ \\ \hline
$124.5$ & $9244$ & $9189$ \\ \hline
$166$ & $9497$ & $9356$ \\ \hline
$249$ & $9842$ & $9839$ \\ \hline
$415$ & $10538$ & $10534$ \\ \hline
\end{tabular}
\end{center}
\caption{The selfconsistent solution at different Higgs masses. The
second column shows the initial configuration (the non-selfconsistent
Sphaleron) and the third column
the final configuration after the self-consistent minimisation. One
can see that the differences are minimal and the total energy is well
approximated by the initial configuration.
\label{selfconstab}}
\end{table}

One should keep in mind that the errors of the selfconsistent calculation
are about $1\%$ due to the inaccuracy of the minimisation process.
However the influence of the self-consistency is negligible, and the
simplest way to compute the one-loop sphaleron energy seems to be to
minimise the classical part of \eqref{esigma} and then add the fermionic
contribution. Of course one has to use the one-loop renormalised 
parameters in evaluating \eqref{esigma} and in computing the final
energy.

\begin{figure}[htb]
\def\epsfsize#1#2{0.7#1}
\centerline{\epsfbox{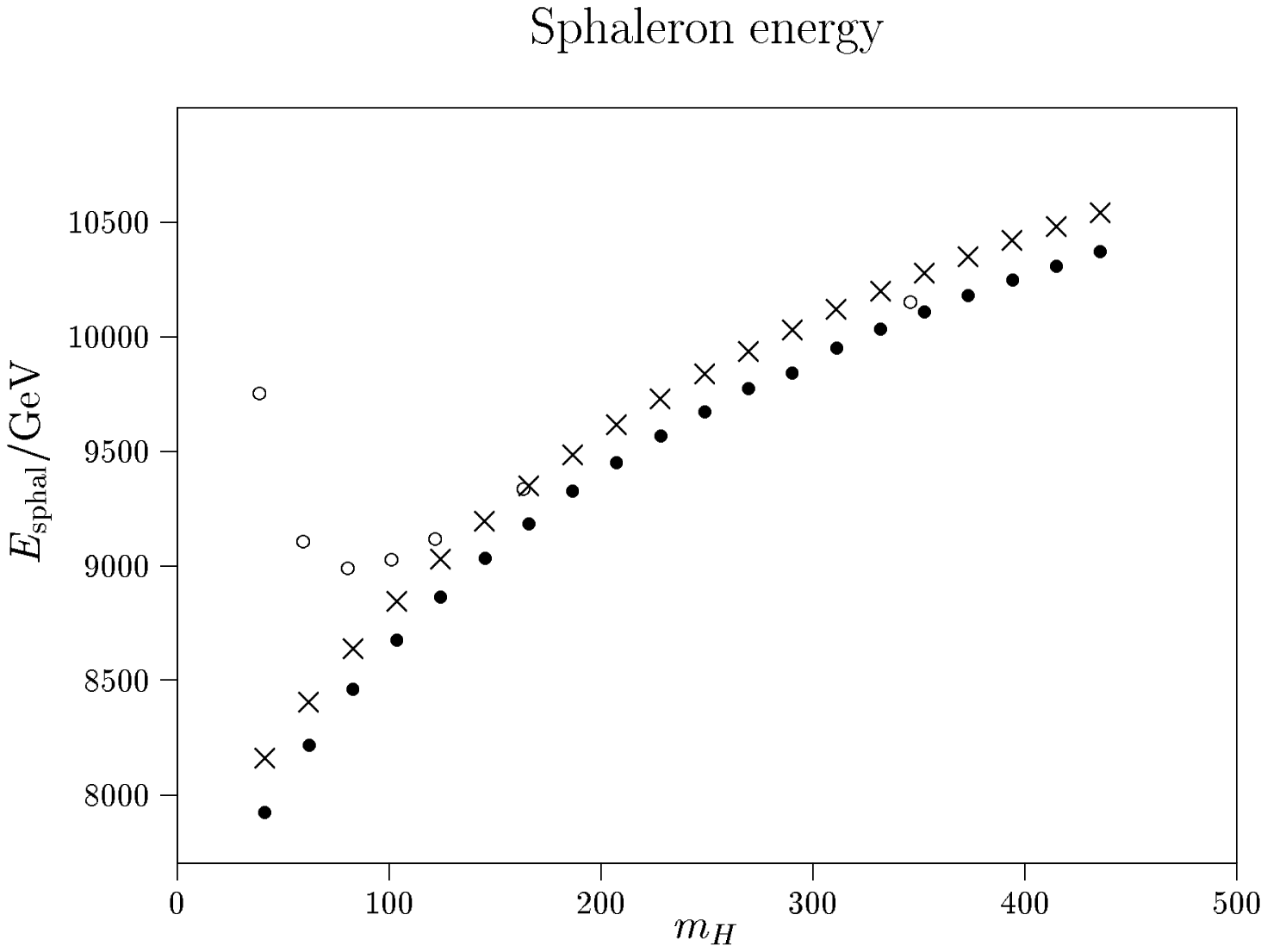}}
\caption{Comparison of the sphaleron energies for different values
of the Higgs mass. The dots represent the energies of the classical
solution, the crosses represent the energies of the one-loop
solutions. The circles represent the results based on the calculations
in \cite{DPSSG} using the renormalization from \cite{sphaleronpaper}.
\label{treeonegraph}}
\end{figure}

With this recipe at hand we can compare the one-loop and the tree-level
energies of the sphaleron for a wider range of Higgs masses. 
This has been done in figure \ref{treeonegraph}. 
One should remember that the
vacuum configuration is no longer stable when the Higgs mass becomes
too small; this phenomenon has also been used to find a lower bound
for the Higgs mass (see e.g. \cite{vacstab}). However, in this work it
only causes that equation \eqref{v1det} cannot be solved numerically 
and we cannot compute the one-loop solution.

It is clearly visible that the shape of the configuration is
maintained in comparison to the classical curve. But the overall energy
has increased by an amount of about $2\%-3\%$. 
From figure \ref{treeonegraph}
we can thus conclude that the influence of the one-loop fluctuations is
only a few percent in comparison to the tree level result.

When comparing these results with the results from previous 
computations \cite{DPSSG,sphaleronpaper}, we find that the previous
results deviate significantly from the current results. The old
curve (represented by circles in \ref{treeonegraph}) is much too
large for small values of the Higgs mass and even exhibits a local
minimum at about $m_H\simeq m_W$. This is a strong indicator for
the incomplete nature of the renormalization scheme used.

%% file: chap4.tex

\section{Conclusion}

In the last section we have found that the correction to the energy
of the sphaleron configuration due to fermionic contributions is very
small, only about $2\%-3\%$. Furthermore in computing this effect the
influence of self-consistency turned out to be negligibly small. 
Since the fermionic
contribution is only small in principle the shape of the dependency
of the sphaleron energy on the Higgs mass did not change - it is only
shifted by a small amount upwards. This puts the results in 
\cite{DPSSG} into doubt, since they found due to the incomplete
renormalization a strong increase in the
sphaleron energy even at zero temperature.

Still, an interesting question is whether the results 
of this paper can be generalised
to the case of non-zero temperature. In this case the correction
generated by the fermion determinant is not much smaller than the
classical sphaleron energy. Although it appears possible that now 
self-consistency becomes important, it has been shown in 
\cite{DPSSG} that the fermionic contribution can be split up
in two parts: the major part has the same shape as the classical energy and
the remaining part can be treated as a small correction. Now one can 
include the first part of the fermionic energy into the classical one
and compute the sphaleron energy again (practically this means one has
to adjust the bare-parameters again and then proceed in a similar
way as in this paper). But it was shown in \cite{sphaleronpaper} that
then the remaining fermionic contribution again has the effect of a
small correction to the overall energy. So in this case the energy
of the whole configuration should not be significantly influenced by
self-consistency. So we argue that also in the case of finite
temperature it is unlikely that self-consistency introduces any
changes in the energy.

However, the results obtained in \cite{sphaleronpaper} appear to 
suffer from the same shortcomings (incomplete renormalization) 
as those in \cite{DPSSG} when it
comes to the absolute value of the energy. Thus the conclusions
drawn for the upper limit of the Higgs mass seem to require a
correction that would lower the value and thus move it closer to 
the one cited by Shaposhnikov in \cite{shaposhnikov}.

\vskip 0.25cm
{\bf Acknowledgement:} The authors wish to thank D.~Diakonov, 
P.~Pobylitsa, M.~Polya\-kov, J.~Schaldach, P.~Sieber and C.~Weiss 
for their valuable input and discussion.

%% file: selfconsa.tex
\section{Mass renormalisation}\label{appendixmass}

Here we want to present the shape of the equations \eqref{massrenorm}.
These were calculated in \cite{docjoerg}.

On tree level we have
\begin{eqnarray}
m_W&=&83\mbox{GeV},\nonumber\\
m_H&=&\nu_H m_W,\nonumber\\
m_F&=&\nu_F m_W.
\label{classmasses}
\end{eqnarray}
 
In the case of one-loop fluctuations we find the following set of
equations to determine the bare parameters:
\begin{eqnarray}
0&\equiv&\displaystyle
-m_H^2+{\nu_H^2\over 8}\left(3v_1^2-4\right)
\nonumber\\&&\displaystyle\quad
+{g^2\over 16\pi^2}{N_c\over 2}\left(
2\nu_F^2\nu_{\rm ren}^2F'\left(
\frac{v_1^2\nu_F^2}{4\nu_{\rm ren}^2}\right)\right.
\nonumber\\&&\displaystyle\quad\quad\left.
\nu_F^2\left(v_1^2\nu_F^2-m_H^2
\left(-C-G\left({m_H^2\over\nu_{\rm ren}^2},
{v_1^2\nu_F^2\over 4\nu_{\rm ren}^2}\right)\right)\right)
\right)\nonumber\\
0&\equiv&\displaystyle
-m_W^2+{v_1^2\over 4}
\nonumber\\&&\displaystyle\quad
+{g^2\over 16\pi^2}\left({N_c\over2}\left(
\nu_{\rm ren}^2F'\left({v_1^2\nu_F^2\over
4\nu_{\rm ren}^2}\right)
+{v_1^2\nu_F^2/2-m_W^2\over 2}\times
\right.\right.
\nonumber\\&&\displaystyle\qquad\quad\times
\left(-C-G\left({m_W^2\over
\nu_{\rm ren}^2},{v_1^2\nu_F^2\over
4\nu_{\rm ren}^2}\right)\right.
-\nu_{\rm ren}^2\left(-1+(1-C)
\left({v_1^2\nu_F^2\over 4\nu_{\rm
ren}^2}-{m_W^2\over 6\nu_{\rm ren}^2}
\right)\right)
\nonumber\\&&\displaystyle\qquad\qquad
-{m_W^2\over\nu_{\rm ren}^2}
G'\left({m_W^2\over\nu_{\rm ren}^2},
{v_1^2\nu_F^2\over 4\nu_{\rm ren}^2}\right)
-\left({v_1^2\nu_F^2\over 4\nu{\rm ren}^2}
-{m_W^2\over 4\nu_{\rm ren}^2}\right)
\times\nonumber\\&&\displaystyle\qquad\qquad%
\quad\left.\times
\left(-C-G\left({m_W^2\over\nu_{\rm
ren}^2},{v_1^2\nu_F^2\over 4\nu_{\rm ren}^2}
\right)\right)\right)
\nonumber\\&&\displaystyle\quad
+N_c\left(N_g+{1\over 2}\right)\left(
-\nu_{\rm ren}^2-{m_W^2\over 2}
\left(-C-G\left({m_W^2\over\nu_{\rm ren}^2}
,0\right)\right)\right.
\nonumber\\&&\displaystyle\quad\quad\left.\left.
\left.-\nu_{\rm ren}^2\left(-1+(C-1)\left(
{m_W^2\over 6\nu_{\rm ren}^2}\right)
-{m_W^2\over\nu_{\rm ren}^2}G'\left(
{m_W^2\over\nu_{\rm ren}^2},0\right)
\right)\right)\right)\right)\nonumber\\
0&\equiv&{1\over 4}\nu_F^2v_1^2-m_F^2,
\label{fermrenmasses}
\end{eqnarray}
where we have used the following definitions:
\[
G\left(x,y\right)=\int_{-1/2}^{1/2}
d\alpha \ln\left(x\left(\alpha^2-{1\over 4}
\right)+y-i\varepsilon\right)
\]\[
G'\left(x,y\right)=\int_{-1/2}^{1/2}
d\alpha\, \alpha^2\ln\left(x\left(\alpha^2
-{1\over 4}\right)+y-i\varepsilon\right).
\]

\cleardoublepage

%% file: nondeg.tex
\section{Nondegenerate Fermion Masses} \label{nondeg}

As mentioned in the main text, it is not possible in our approach to treat
nondegenerate fermion masses exactly. So we introduced a common fermion mass
$\nuF$ and a degenerate mass matrix
\begin{equation}
M_0=\frac{1}{2}\ \nu_{\rm F}\ \left( 
    \begin{array}{cc}
        \Phi^{0*} & \Phi^+ \\
       -\Phi^{+*} & \Phi^0
     \end{array} \right)
\end{equation} 
to compute the results presented here. In this 
appendix we try to estimate the error of that
approximation by treating nondegenerate fermion masses perturbatively.
In order to restore the original mass matrix $M$, (\ref{massmat}), one can
write 
\begin{equation}
M=M_0\left(1+\frac{\Delta\nu}{\nu_{\rm F}}\tau_3\right),
\end{equation}
where we introduced the parameters 
$\DS \nu_{\rm F}=\frac{\nu_u + \nu_d}{2}$ and $\DS
\Delta\nu= \frac{\nu_u - \nu_d}{2}$.
One can see that the matrix $\tau_3$ spoils the spherical symmetry of our
fermionic Hamiltonian \eqref{hamilton} so that a direct diagonalisation can
not be performed. But
for small mass differences one can now expand $M$ in a power series with
respect to $\Delta\nu / \nuF$. Of course, in nature this parameter for the
massive fermion doublet is
of order $O(1)$, so it is clear, that it is not possible to compute good
estimates with this ansatz. On the other hand, if we are only interested
in the qualitative behaviour of the fermionic energy, it can be used to see
how the 
energy changes if one allows different fermion masses within one doublet.
The fermionic 1-loop energy can then be evaluated as a Taylor series in the
parameter $\Delta\nu$:
\begin{equation} \label{taylor}
E_{\rm Ferm}(\Delta\nu)=E_{\rm Ferm}(0)+\left.\frac{\partial 
    E_{\rm Ferm}}{\partial \Delta\nu}\right\vert_{\Delta\nu=0} \!\!\!\cdot
    \Delta\nu  + \left.\frac{1}{2}\frac{\partial^2  E_{\rm Ferm}}{
   \partial \Delta\nu^2}
   \right\vert_{\Delta\nu=0} \!\!\!\cdot \Delta\nu^2 + O(\Delta\nu^3).
\end{equation}
In this series the linear term vanishes, because in a pure $SU(2)$ gauge
theory up- and down-components in one doublet cannot be distinguished if they
are degenerate in mass. Therefore we compute numerically the first and third
term of this series. To compare our result 
to the approximation we proceed in the
following way: We start with degenerate fermion masses with $\nuF=\nut$ and
count only $n= 3/2$ massive fermion doublets, so that in the end three top quarks
masses are counted. This is exactly the approximation
which was used before. Then we introduce small mass differences $\Delta\nu$,
but keep the value of $\nut$ fixed. To ensure that we have the correct ``mass
content'' we count 
\begin{equation}
n=\frac{3}{2} \frac{\nut}{\nut- \Delta \nu}
\end{equation}  
massive fermion doublets, in order to preserve the sum of three top quark
masses in the fermionic energy.

\begin{figure}[hb]
\epsfxsize 2.7 in
\centerline{\epsfbox[200 450 400 730]{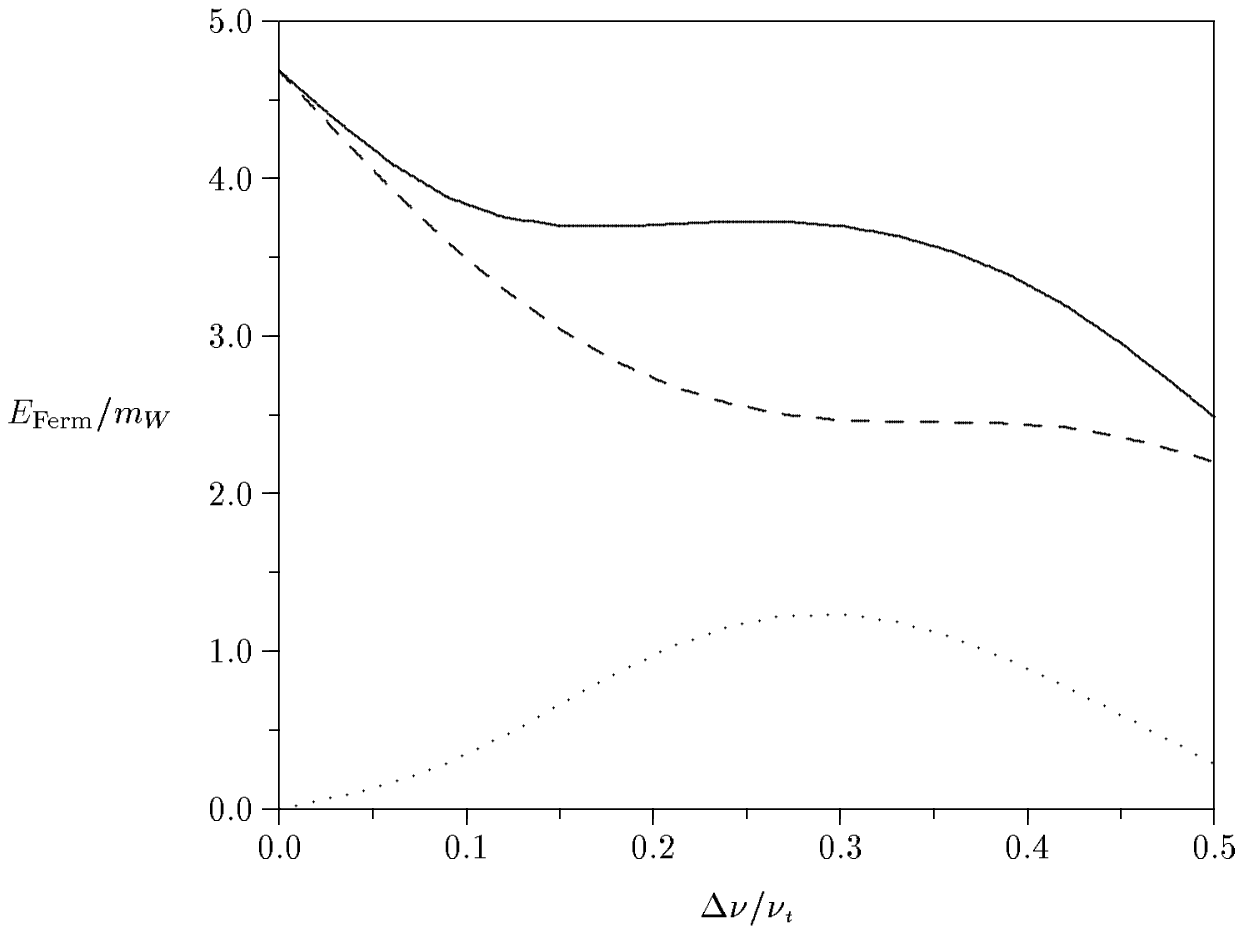}}
\caption{\label{deltam} In this picture one can see the fermionic energy as a
  function of the mass splitting $\Delta\nu$. The dashed line is the first
  term of the series \eqref{taylor} and the dotted line shows the third
  term. The solid line is the sum of the two contributions and therefore the
  whole fermionic energy up to $\Delta\nu^4$, since all odd powers vanish. One
  can see that even a small mass difference lowers the fermionic energy
  significantly.}
\end{figure}

By increasing $\Delta\nu$ we eventually reach the ``physical'' point with
$\Delta\nu=\nut/2$ and $n=3$. This is plotted in figure \ref{deltam}. 
This method will work well in the region of small $\Delta\nu$, since
perturbation theory is well justified here.
We can clearly see, how the fermionic energy drops if one introduces
mass splittings. From this point of view we can at least say that our
approximation is a good upper limit for the energy arising from fermionic
fluctuations.  

\cleardoublepage